\documentstyle[epsfig,float,twocolumn,prb,aps]{revtex}

\begin{document}
\draft

 \twocolumn[\hsize\textwidth\columnwidth\hsize\csname @twocolumnfalse\endcsname

\title{
Logarithmic corrections from ferromagnetic impurity ending bonds  of open
antiferromagnetic host chains 
}

\author{
Jizhong Lou${}^{1}$, 
Jianhui Dai$^{2}$
Shaojin Qin$^{1}$, 
Zhaobin Su$^{1}$,
and Lu Yu$^{3,1}$
}

\address{
$^{1}$Institute of Theoretical Physics, P O Box 2735, Beijing 100080,
        P R China\\
$^2$Zhejiang Institute of Modern Physics, Zhejiang University, Hangzhou 310027, 
		P R China\\
$^3$International Center for Theoretical Physics, P. O. Box 586, 
	34100 Trieste, Italy\\
}

\date{\today}  
\maketitle
\begin{abstract}
We analyze the logarithmic corrections due to  ferromagnetic impurity  
  ending bonds of open spin 1/2 antiferromagnetic chains, using the density matrix renormalization
group technique. A  universal finite size 
scaling $\sim {\frac 1 {L \log L }}$ for  impurity contributions in the quasi-degenerate
  ground state  energy is demonstrated for a zigzag spin  1/2 
chain  at the critical next nearest neighbor coupling  and the standard Heisenberg spin  1/2  chain, in the
long chain   limit.  Using an exact solution for the latter case it is argued that one can
extract the impurity contributions to the entropy and specific heat from the scaling analysis.
It is also shown that a pure spin $3/2$ open Heisenberg  chain belongs to the same
universality class.
\end{abstract}

 ]

The logarithmic corrections due to marginally irrelevant operators\cite{cardy}
 complicate greatly the comparison of experiments and numerical simulation
results for finite size systems  with analytical calculations.
However, a careful analysis in some specific cases may yield useful information
on the low energy excitation spectrum and relevant physical quantities.
In this Report, we consider the finite size scaling for open spin 1/2
antiferromagnetic (AF) chains with ferromagnetic (FM) coupling at the ending bonds. 
This system exhibits the same behavior as a  Kondo 
 impurity coupled ferromagnetically to a Luttinger liquid.\cite{k2,k4,k5,ypw}
 In these systems the Kondo screening is not complete, and the ground state is quasi-degenerate, 
{\it i.e.}  the level spacing is vanishing faster than $1/L$, where $L$ is the system size.
 Recently, some exactly solvable models  belonging to 
this class  have been found, and the impurity entropy as well as
 specific heat has been obtained  through thermodynamic Bethe ansatz\cite{k5,ypw,dwe}. 
We will show in this report the universal behavior of logarithmic corrections 
for this class of systems and, using the exact solution,  we will argue
that one can extract from the scaling analysis the impurity contributions to the entropy
and specific heat  when the exact solutions are not available.
We will also show that a pure open spin 3/2  AF chain (without additional FM
ending bonds) belongs to the same universality class.

\begin{figure}[ht]
  \epsfxsize=3.3 in\centerline{\epsffile{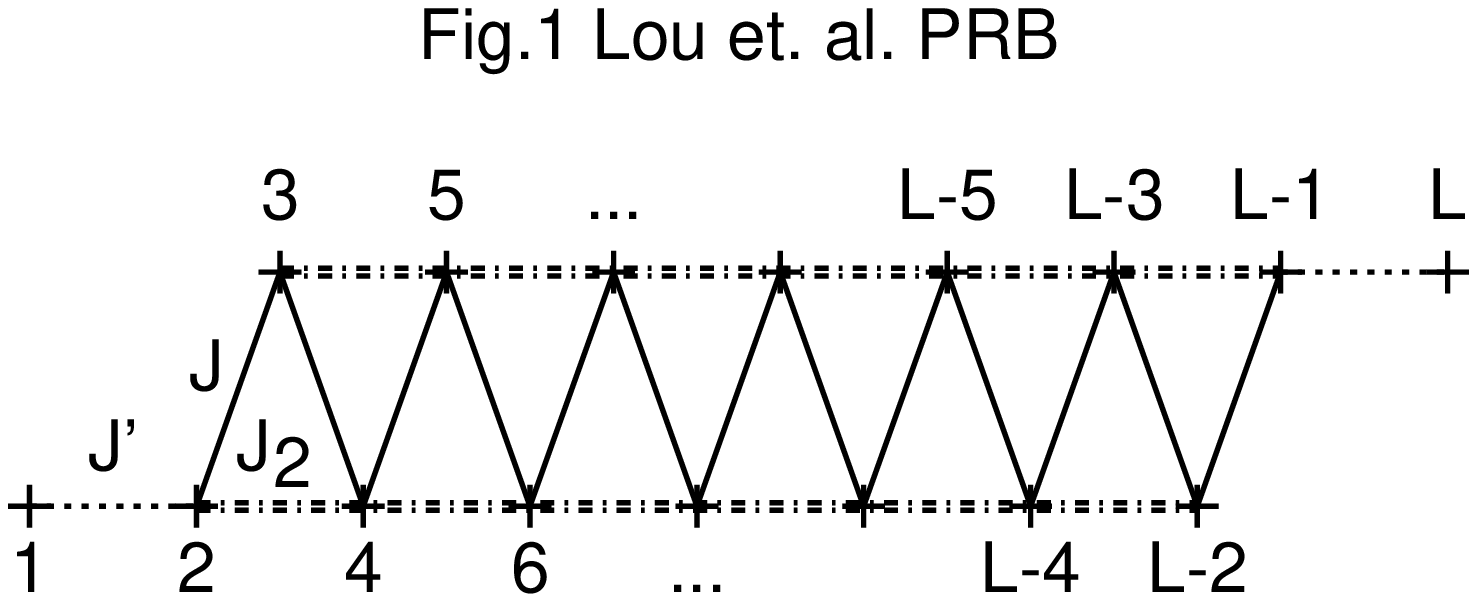}}
\vspace{0.5cm}
\caption[]{
Open  zigzag spin chain with coupling $J$ and $J_2$ for nearest 
  and next nearest neighbors, respectively. Two ending impurity 
spins couple to the bulk by  a FM bond $J'<0$.
}
\label{figzg}
\end{figure}

The impurity effects in  spin 1/2   Heisenberg AF chains have 
 been discussed  extensively in Ref.\onlinecite{egg}. 
The logarithmic corrections  to scaling functions for spin 1/2 Heisenberg
chains have been discussed in recent papers.\cite{bar,aff1,bru,tsai}
Those corrections are due to bulk marginally irrelevant
operators, although their manifestations depend on boundary conditions.\cite{aff1}
Instead,  we will carry out   a  detailed numerical analysis of 
the finite size scaling  for the ground state near-degeneracy and low energy 
spectrum of  open  $s=1/2$ spin chains
due to FM ending bonds. To the best of our knowledge, this issue has not been  addressed
numerically up to now. We  first consider the following Heisenberg
chain with next nearest neighbor coupling, or equivalently, a zigzag
chain:
\begin{eqnarray}
H & = & \sum_{i=2}^{L-2}{\bf S}_{i}\cdot{\bf S}_{i+1} 
	+J_{2c} \sum_{i=2}^{L-3}{\bf S}_{i}\cdot{\bf S}_{i+2} 
	+H_{imp}, \nonumber \\
H_{imp} &=& -{\bf S}_{1}\cdot{\bf S}_{2}-{\bf S}_{L-1}\cdot{\bf S}_{L},
\label{hamj12}
\end{eqnarray}
where $L$ is the chain length and ${\bf S}_{i}$ is the $s=1/2$ spin on 
site $i$.  We draw the system in Fig. \ref{figzg}.  The nearest 
neighbor coupling is set to $J=1$ and the next nearest neighbor 
coupling is set to the critical value\cite{oka} $J_2=J_{2c}=0.2411$.  
The two ending bonds are FM $J'=-1$, and the ending sites 
are impurity  spins, also $s = 1/2$.   There are
no logarithmic corrections due to zero initial bulk 
marginal coupling at the critical value $J_2=J_{2c}$.\cite{hal,aff}   Therefore,
all logarithmic corrections in our calculations are coming entirely from the boundary
effects.

A pure open zigzag chain without $H_{imp}$ has the same low energy 
spectrum as the pure open $s=1/2$ chain when $0<J_2<J_{2c}$.  
 There is  a unique ground state with total 
spin   $S=0$,  and one first excited state of spin   $S=1$, with 
excitation energy scaled as $\pi v /L$ for finite size chains,\cite{egg}
where $v$ is the spin velocity.
For the zigzag chain shown in Fig. \ref{figzg} with FM ending bonds,
 the ending spins are not fully  screened, and there is an RKKY coupling 
between them scaled as $J_{RKKY}(L) = {\frac a {L \log L }}$ in the large $L$ limit.
The two ending impurity spins  form a   singlet and a  triplet with energy spacing $J_{RKKY}(L)$. 
We   identify the following  low energy states:\\
(1) Two quasi-degenerate ground states: One is  composed of  the bulk $S=0$ 
	state and singlet impurity state with energy  $E_0^{O}$;  the other is 
	formed by  the bulk $S=0$ state and  the triplet impurity state 
with  energy 	$E_1^{O}$.  We take  $J_{RKKY}(L)=E_1^{O}-E_0^{O}$ 
 as the definition of $J_{RKKY}(L)$. \\
(2) Four quasi-degenerate first excited states with excitation energy scaled 
      as $\pi v /L$.  One is composed of  the bulk
	$S=1$ state and singlet impurity state with  energy $E_1^{I}$.  
The other  three are formed by the bulk first excited   $S=1$ state 
	and  triplet impurity state with energies $E_2^{II}$, $E_1^{II}$, and 
	$E_0^{II}$,  for the total spin $S=2, 1$, and 0, respectively. 
	Due to the  bulk $S=1$ excitation propagating in-between the ending spins, 
the energy difference $E_i^{II}-E_j^{II}$ is not the same as $J_{RKKY}(L)$.

\begin{figure}[ht]
  \epsfxsize=3.3 in\centerline{\epsffile{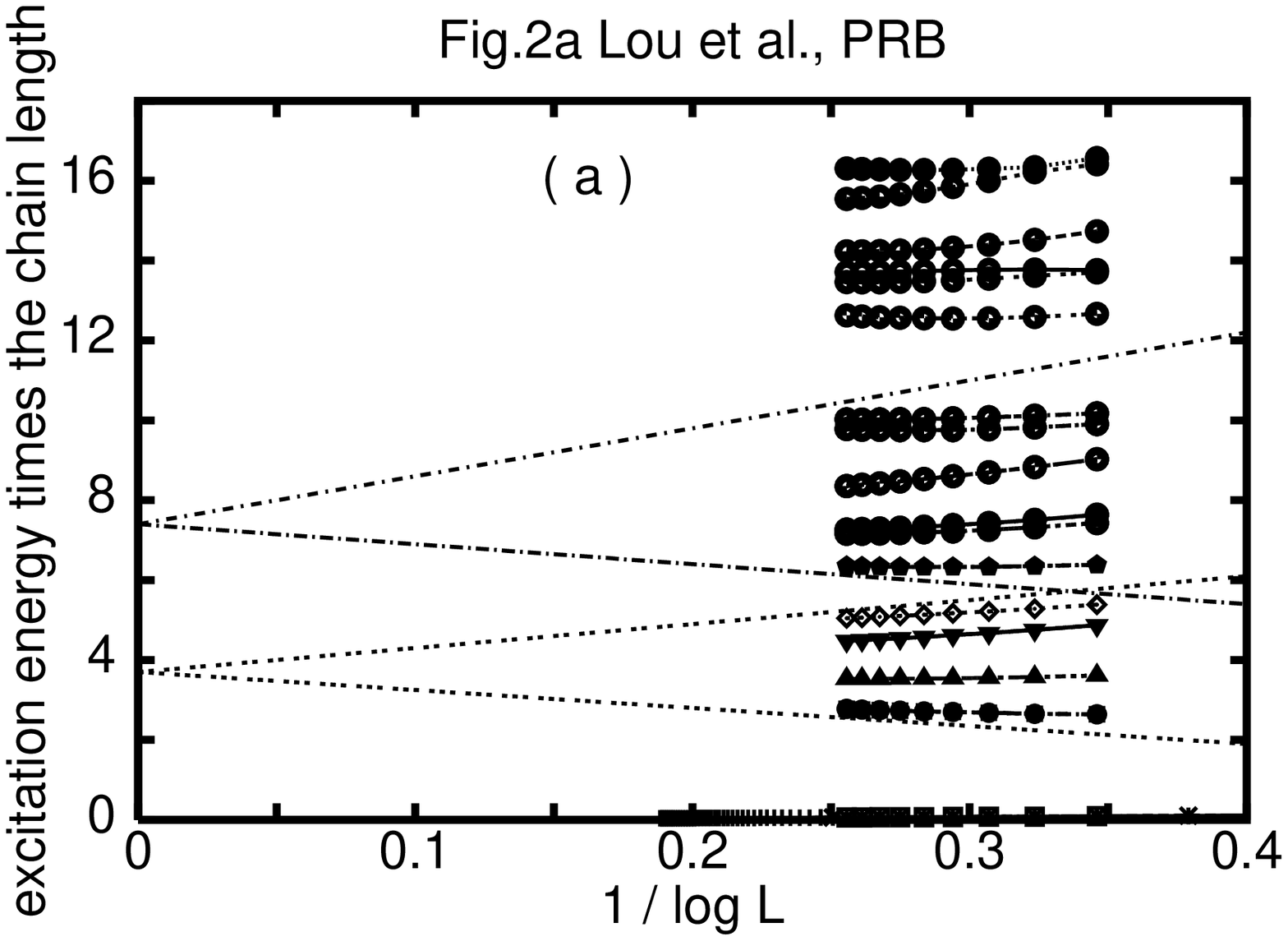}}
  \epsfxsize=3.3 in\centerline{\epsffile{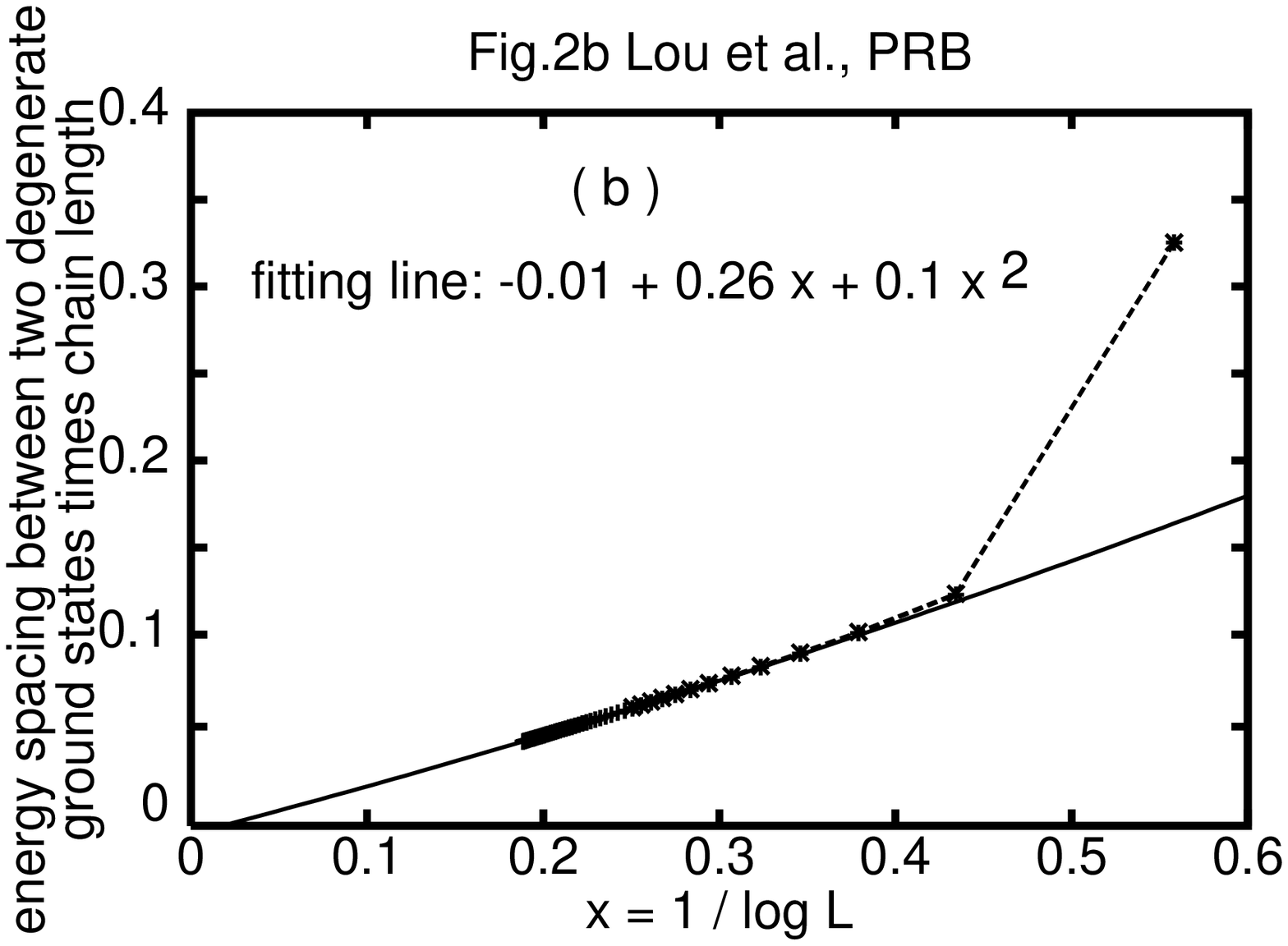}}
\vspace{0.5cm}
\caption[]{
Excitation energies times the chain length, $(E_i^{K}-E_0^{O})L$,
vs $1/\log L$ are  plotted in figure (a) for the zigzag chain shown in 
Fig. \ref{figzg}.  Scaling for the quasi-degenerate ground state, 
$(E_1^{O}-E_0^{O})L$ vs $1/\log L$, is plotted in figure (b) along with a 
fitting line.
}
\label{enzg}
\end{figure}

We use density matrix renormalization group (DMRG)  method\cite{wht}   to 
calculate low energy levels for the above Hamiltonian.  By keeping 
$m=150$ states, the truncation error is as small as $10^{-7}$.
We study even length chains only. The low-lying excitation energies
 are plotted vs $1/\log L $ in Fig. \ref{enzg}.  In 
Fig. \ref{enzg}a we see  the ground state is degenerate at the scale
of the graph. The first four 
excitation energies scale  to $\pi v /L$.  
They correspond to
$E_2^{II}$, $E_1^{I}$, $E_1^{II}$, and $E_0^{II}$, respectively,   from bottom up.  We note  the ratio between the two energy 
spacings $E_0^{II}-E_1^{II}$ and $E_1^{II}-E_2^{II}$  is approximately 
$1:2$, which indicates 
 these excitations can be identified as due to   coupling of the bulk $S=1$ state 
and impurity triplet state. The next group of six energy levels 
scales to $2\pi v/L$.  They are   composed of the impurity 
singlet/triplet states and  the two bulk levels\cite{egg} scaled as 
$2\pi v/L$ of spin $S=0$ and $S=1$.  We have drawn guiding lines to 
separate these two groups of lowest excitations.
In Fig. \ref{enzg}b, we magnify the scale of the energy spacing for the quasi-degenerate
 ground states to show the scaling 
$E_1^{O}-E_0^{O} =J_{RKKY}(L) \sim {\frac 1 {L \log L}}$. 
Since there are no bulk logarithmic corrections involved at the critical 
coupling $J_{2c}$ for zigzag  chains, the logarithmic term appears 
solely due to the Kondo impurity effect. 
If the ending bound coupling $J'=0$, the end  impurity spins are  decoupled from 
the bulk and the ground state  is exactly degenerate.
 The impurity has nonzero entropy at zero 
temperature.  When $J'<0$, the ground state and low energy spectrum 
have an asymptotic degeneracy and the energy difference between these 
quasi-degenerate  states scales as ${\frac 1 {L \log L}}$.  This
has been very clearly seen  for the ground state. The zero temperature
entropy  will change, as we shall argue, depending on   the coefficient 
$a$ in the ground state energy 
scaling $J_{RKKY}(L)=  {\frac a {L \log L}}$.  
 We will see that this logarithmic scaling   behavior  due to boundary spins is universal for  
  systems composed of  Kondo   impurities ferromagnetically coupled  to Luttinger liquids,
even when there are also logarithmic corrections due to bulk marginally irrelevant operators.

\begin{figure}[ht]
  \epsfxsize=3.3 in\centerline{\epsffile{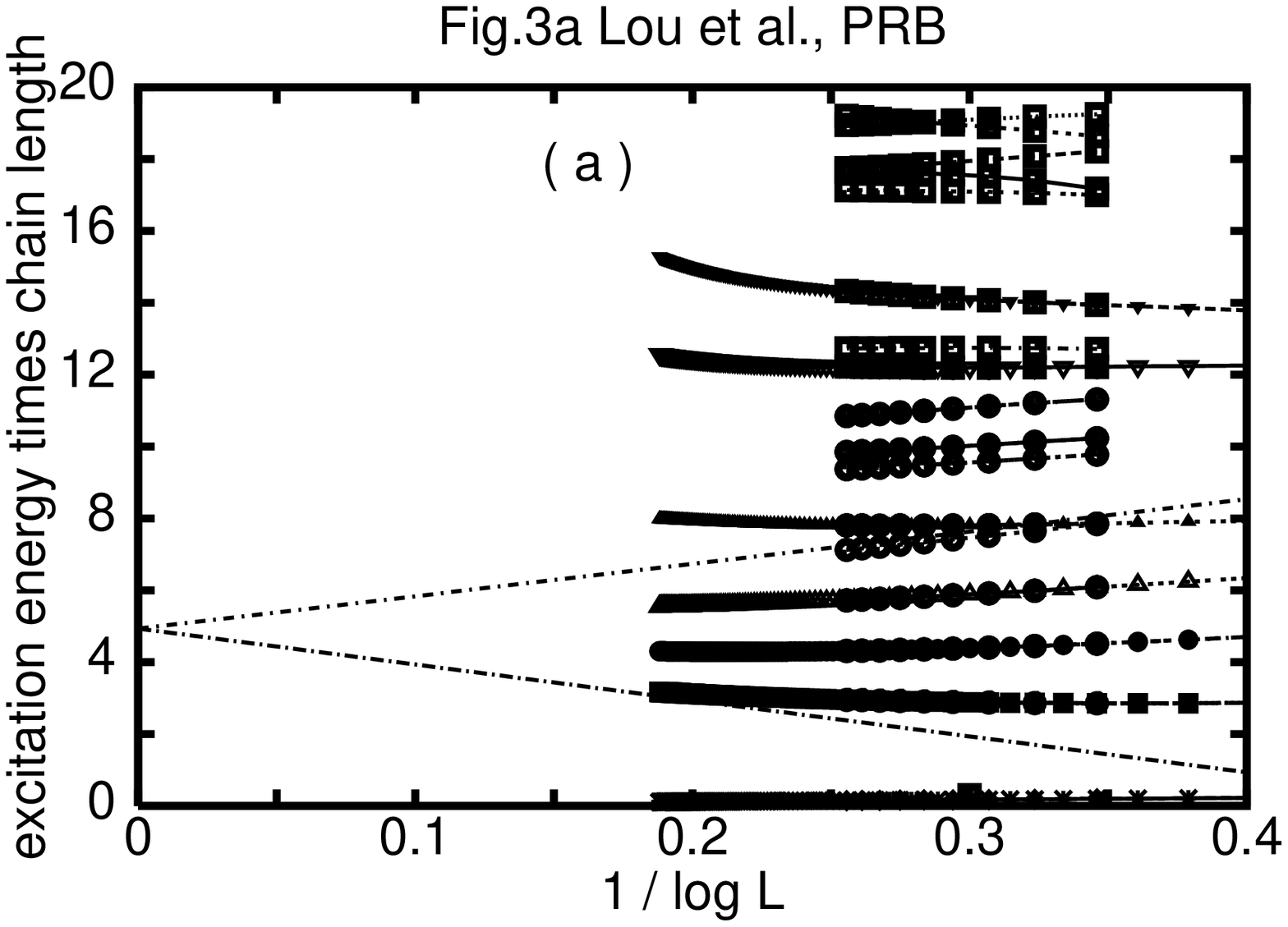}}
  \epsfxsize=3.3 in\centerline{\epsffile{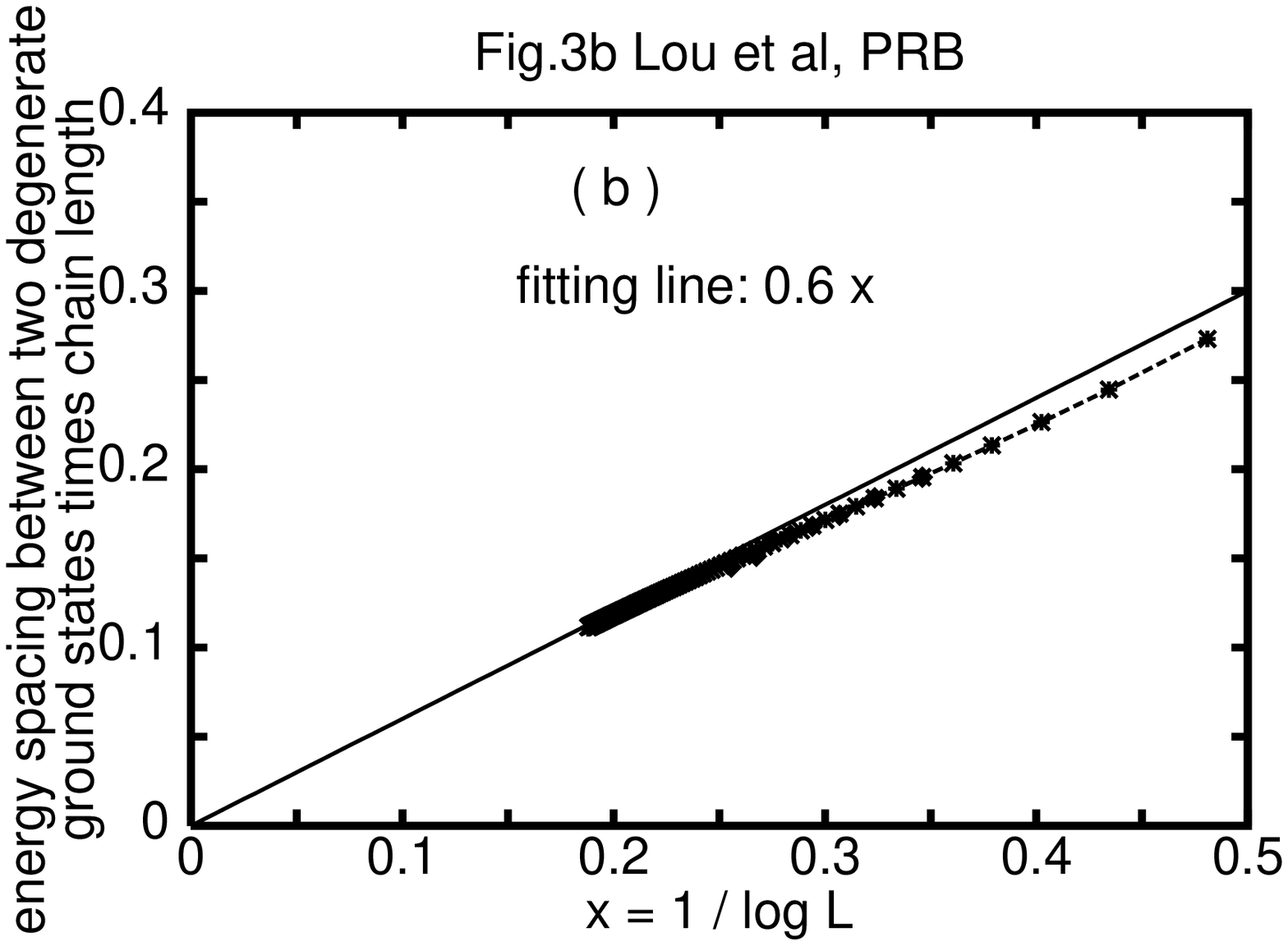}}
\vspace{0.5cm}
\caption[]{
Excitation energies times the chain length, $(E_i^{K}-E_0^{O})L$, vs 
$1/\log L$ are plotted in figure (a) for spin 1/2 Heisenberg AF chain
with FM ending bonds.  Scaling for the quasi-degenerate ground state, 
$(E_1^{O}-E_0^{O})L$ vs $1/\log L$, is plotted in figure (b)  along with a 
fitting line.
}
\label{ens05}
\end{figure}

We consider now an open spin 1/2 AF Heisenberg chain 
with impurity ending bonds described by the  Hamiltonian 
\begin{eqnarray}
H & = & \sum_{i=2}^{L-2}{\bf S}_{i}\cdot{\bf S}_{i+1} 
-{\bf S}_{1}\cdot{\bf S}_{2}-{\bf S}_{L-1}\cdot{\bf S}_{L}.
\label{hams05}
\end{eqnarray}
The nearest neighbor  coupling  is set to $J=1$ and the FM  
coupling $J'$ for the ending spin is set to $J'=-1$. For a pure spin 1/2 
open chain without    FM impurity bonds, the ground 
state energy scales as\cite{aff1}
$E = e_0 L + e_1 - {\frac {\pi v}{24 L}}[ 1 + b/\log^2(L) +\ldots]$, 
where   $e_0$ is the site energy, $e_1$ is the boundary energy, and 
  $v$ is the spin velocity for spin 1/2 Heisenberg chain.  The logarithmic
correction appears here due to the bulk marginally irrelevant operator. We 
will demonstrate the ground state energy 
has one more term 
${\frac 1 {L \log L}}$ in its finite size scaling due to FM 
Kondo coupling:
\begin{eqnarray}
E_1^{O}=e_0 L + e_1 - 
{\frac {\pi v} {24 L}}  + {\frac a {L\log L}} +\ldots.
\label{egs1}
\end{eqnarray}
  We calculate 
the energy levels by using DMRG method for even length chains.  We keep 
$m=200$ states and the truncation error is  of the order  $10^{-9}$.  The 
ground state is the same   as for Hamiltonian 
(\ref{hamj12}),  with energy $E_0^{O}$.  The energy levels $E_i^{K}$ are also labeled  the 
same way.  We plot  the excitation energies 
  $(E_i^{K}-E_0^{O})L$ vs $1/\log L$ in Fig. \ref{ens05}a.  
In  Fig.3b, the scaling $E_1^{O}-E_0^{O}\sim {\frac {0.6} {L \log L}}$ 
is exhibited.  The first excited states are four-fold degenerate as 
we analyzed before  for the zigzag chains.  We have drawn guiding lines in 
Fig. \ref{ens05} to group these first excited states together.  For the low 
energy spectrum, the excitations are again composed of the  combined impurity 
and  bulk spin states. The logarithmic scaling behavior of  a standard 
Heisenberg chain  due to FM impurity bonds
is identical to that of zigzag chain, as the correction due to the bulk marginally irrelevant operator
is of higher order.

\begin{figure}[ht]
  \epsfxsize=3.3 in\centerline{\epsffile{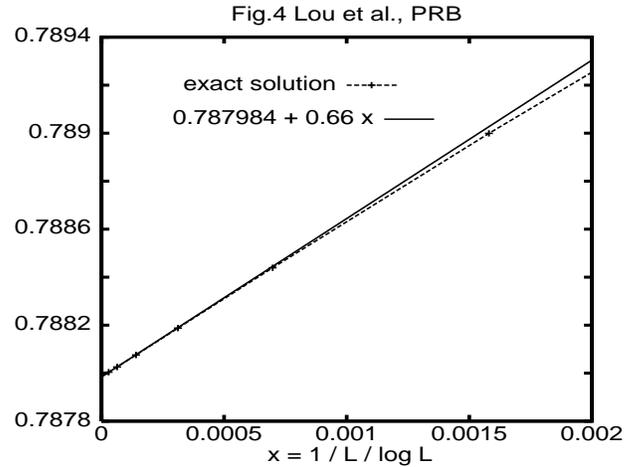}}
\vspace{0.5cm}
\caption[]{
Energy of one of the  quasi-degenerate ground states with total spin $S=1$
$E_1^{O}$, obtained from the exact solution\cite{ypw} is plotted 
vs $ { 1 \over {L \log L}}$  along with the fitting line.   
}
\label{enba}
\end{figure}

On the other hand,  the model(\ref{hams05}) has been solved exactly 
using the Bethe Ansatz,\cite{ypw}
we  can extract  its ground state energy from the exact solution.
 We calculate, using the Bethe Ansatz equations of Ref. \onlinecite{ypw}, the energy 
$E_1^{O}$ of $S = 1$ state for system size up to more than four thousands
sites.\cite{sore}  (We have not yet  obtained the energy for another 
degenerate ground state with lower energy $E_0^{O}$ from the Bethe 
ansatz equations.)  Following Eq.(\ref{egs1}), we 
plot $E_1^{O} - e_0 L + {\frac {\pi v} {24 L}}$ vs 
${\frac 1 {L\log L}}$ in Fig. \ref{enba}, where the site energy 
$e_0 = 1/4 - \log {2}$, and the spin velocity  $v=\pi/2$.  We obtain 
$E_1^{O} - e_0 L + {\frac {\pi v} {24 L}} 
= 0.787984 + {\frac {0.66} {L\log L}}$ by the least square fitting.  
Based on  Eq.(\ref{egs1}), we have $e_1=0.787984$.  We obtain 
also the scaling $J_{RKKY}(L)\sim  {\frac {0.66} {L \log L}}$ from the 
exact solution, to be compared with an almost identical 
fitting of the DMRG data  (coefficient  $a = 0.6$ instead of 0.66 in the
scaling formula $J_{RKKY}(L)=  {\frac a {L \log L}}$). Therefore, we 
conclude that the quasi-degenerate ground state and low-lying excitations
of Luttinger liquids with FM Kondo impurities 
can be indeed described  by this logarithmic scaling function.

Moreover, using the exact solution,\cite{ypw} the impurity entropy and 
specific heat can be expressed  in terms of  the coefficient $a$ 
in the above finite size scaling.  If two systems have the same ground state
degeneracy and the same low-energy excitation spectrum, the thermodynamic
properties at very low temperatures should also be identical to each other. Using this argument,
we can estimate the impurity entropy and specific heat from the finite size 
scaling analysis of the energy spectrum even in the case when the exact solution
is not available ({\it e.g.} the zigzag chain  considered earlier).
 Using the same argument we will discuss another related
system, an open  spin 3/2 AF chain without additional FM ending bonds.

\begin{figure}[ht]
  \epsfxsize=3.3 in\centerline{\epsffile{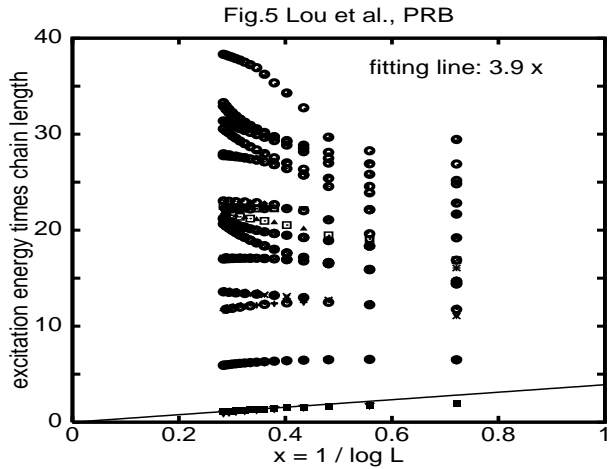}}
\vspace{0.5cm}
\caption[]{
Excitation energies times chain length are  plotted vs $1/\log L$ for 
 open $s=3/2$ chains.   The energy times chain length 
difference between the quasi-degenerate ground states is fit by $3.9/\log L$.
}
\label{ens15}
\end{figure}

The spin $3/2$ chain has been shown to have the same low energy physics as 
for spin 1/2  chain.\cite{halb}  When the chain is open, there are 
effective edge $s'=1/2$ spins left at the ends of the 
chain.\cite{ng} It was   shown earlier\cite{qin} that the energy spacing between 
the two degenerate ground states scales as ${\frac 1 {L \log L}}$.  
However,  the scaling of the low energy spectrum is very difficult to study 
and the logarithmic corrections are big.\cite{halb}  
  Following the way of  dealing with such edge spins for integer 
spin chains,\cite{sor1} it was proposed to treat them as impurity 
spins {\it ferromagnetically}  coupled to the bulk spin excitations.\cite{qin}  
Unlike the previous cases, there are no additional FM impurity bonds
at the ends of spin 3/2 chains. 
 Now we calculate the scaling of the ground state energy spacing 
$J_{RKKY}(L)=  {\frac a {L \log L}}$ by DMRG.  We keep $m=1000$ states 
in DMRG and the truncation error is $10^{-6}$.  We calculate only a 
few low energy levels and plot the excitation  energies times length 
vs $1/\log L$ in Fig. \ref{ens15}.  We obtain the coefficient 
$a=3.9$ in scaling  $J_{RKKY}(L)=  {\frac a {L \log L}}$.  With the 
known spin velocity $v=3.87$ for spin 3/2 chain,\cite{halb} we 
 obtain a  nonzero entropy at zero temperature. Its exact 
value can be expressed in term of $a$ analytically when exact solution is available.(Non-zero impurity entropy is also predicted in Takhatajan-Babujian spin-3/2 chain with spin-1/2 boundary impurities, where Bethe ansatz solution is available.\cite{dwe})
 We hope 
such a  nonzero entropy at zero temperature can be measured in quasi-
one-dimensional spin 3/2 materials such as  $CsVCl_3$ (Ref. \onlinecite{e1}), 
$AgCrP_2S_6$ (Ref. \onlinecite{e2}), etc. 

In summary, we have carried out a detailed analysis of  the low energy spectrum
 of  FM Kondo impurity in Luttinger liquids, or equivalently, an open  spin
1/2 AF Heisenberg chain.  We have shown   this class of models has a
universal logarithmic quasi-degeneracy in the low energy states due to the
RKKY interaction between the unscreened edge spins.
We argue  that nonzero  entropy at zero temperature can be 
obtained for  FM Kondo 
impurity system by studying the finite size scaling of the ground state 
energies.  

J. Lou and S. Qin would like to thank Prof.T.K. Ng for valuable discussions.  
This work is partially supported by Chinese Natural Science Foundation.

\end{document}